# A Novel Spatiotemporal Correlation Anomaly Detection Method Based on Time-Frequency-Domain Feature Fusion and a Dynamic Graph Neural Network in Wireless Sensor Network

Miao Ye, Zhibang Jiang, Xingsi Xue, *Member, IEEE*, Xingwang Li, *Senior Member, IEEE*, Peng Wen, Yong Wang

*Abstract*—Attention-based transformers have played an important role in wireless sensor network (WSN) timing anomaly detection due to their ability to capture long-term dependencies. However, there are several issues that must be addressed, such as the fact that their ability to capture long-term dependencies is not completely reliable, their computational complexity levels are high, and the spatiotemporal features of WSN timing data are not sufficiently extracted for detecting the correlation anomalies of multinode WSN timing data. To address these limitations, this paper proposes a WSN anomaly detection method that integrates frequency-domain features with dynamic graph neural networks (GNN) under a designed self-encoder reconstruction framework. First, the discrete wavelet transform effectively decomposes trend and seasonal components of time series to solve the poor long-term reliability of transformers. Second, a frequency-domain attention mechanism is designed to make full use of the difference between the amplitude distributions of normal data and anomalous data in this domain. Finally, a multimodal fusion-based dynamic graph convolutional network (MFDGCN) is designed by combining an attention mechanism and a graph convolutional network (GCN) to adaptively extract spatial correlation features. A series of experiments conducted on public datasets and their results demonstrate that the anomaly detection method designed in this paper exhibits superior precision and recall than the existing methods do, with an F1 score of 93.5%, representing an improvement of 2.9% over that of the existing models. Code is publicly available at https://github.com/GuetYe/anomaly_detection.

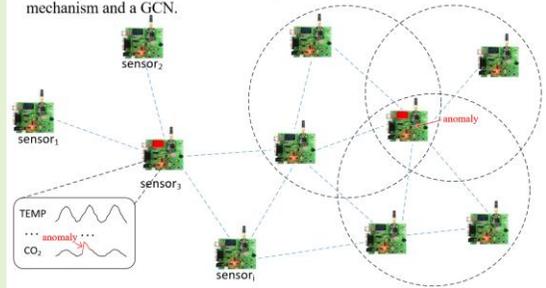

## I. INTRODUCTION

WIRELESS sensor networks (WSN) are multihop self-organizing networks that are composed of numerous independently distributed sensor nodes. Each node has the ability to independently collect physical information with various modalities, such as temperature and humidity information, and can be used to monitor the environmental conditions of the deployment area in real time. These sensor nodes are interconnected through wireless communication, offering flexible and convenient deployment opportunities. WSN are widely used in various fields, including national defense and the military[1], natural environment monitoring[2], medical surveillance[3], agricultural production monitoring[4], and smart cities[5].

The sensor nodes in WSN are often located in complex environments such as the wild, deserts, or forests, making them susceptible to natural disasters, environmental interference, human intrusion, or device malfunctions[6]. These factors can cause anomalies in the raw data streams collected by the sensor nodes, resulting in data distortion and affecting the stability and reliability of the utilized system. Designing effective anomaly detection methods is crucial for ensuring the reliability and stability of network operations. Efficiently detecting and localizing anomalies is vital for repairing or eliminating faulty nodes and ensuring the stable operation of WSN[7-9].

Manuscript received August ##, 2024. This work was supported in part by National Natural Science Foundation of China (Nos.62161006, 62172095, 62372353), the subsidization of Innovation Project of Guangxi Graduate Education (No. YCSW2022270), and Guangxi Key Laboratory of Wireless Wideband Communication and Signal Processing (No. GXKL06220110﹐ GXKL06230102)﹐ Ministry of Education Key Laboratory of Cognitive Radio and Information Processing (No. CRKL220103). (Corresponding author: Xingsi Xue)
Miao Ye, Zhibang Jiang, Peng Wen, Yong Wang are with School of Information and Communication, Guilin University of Electronic Technology, Guilin 541004, China(e-mail: ym@mail.xidian.ediu.cn; jiangzhibang33@foxmail.com; wenpeng1994325@gmail.com; ywang@guet.edu.cn).
Xingsi Xue is with Fujian Provincial Key Laboratory of Big Data Mining and Applications, Fujian University of Technology, Fuzhou, Fujian, 350118, China (e-mail: jack8375@gmail.com).
Xingwang Li is with the School of Physics and Electronic Information Engineering, Henan Polytechnic University, China (e-mail: lixingwangbupt@gmail.com).

Typically, sensor nodes collect data concerning multiple physical quantities (such as temperature and humidity) at equal time intervals, forming multitime series data (also referred to as multimodal data in some studies)[10-12]. Anomalies in these WSN multitime series refer to data samples that deviate from the normal data distribution. Anomalies can take the form of point anomalies, contextual anomalies, and collective anomalies[13-16]. Point anomalies are single or few data points that significantly deviate from the remainder of the data. Contextual anomalies are data points that deviate from most normal data points within a specific context. Collective anomalies are anomalies consisting of multiple groups of data points over continuous time intervals relative to the entire dataset; each individual point may not be an anomaly, but their collective occurrence is abnormal.

Additionally, correlations are observed between the multitime series data collected by sensor nodes and the time series data collected by different sensor nodes in a WSN. For example, during a fire, the temperature readings collected by a sensor node increase, whereas the humidity readings decrease, indicating a negative correlation between temperature and humidity. Another example is that during a fire, the temperature readings collected by different sensor nodes around the fire source should all increase, indicating positive correlations among the temperature time series data of these different nodes. Anomalies that violate these normal correlation patterns can be termed "correlation anomalies"[10]. Such correlation anomalies may not be apparent when looking at a single time series but can be detected by comparing related time series.

The traditional WSN anomaly detection methods include signal processing and machine learning techniques. Examples include graph signal filtering [17], subgraph partitioning[18], clustering[19], and tree models[20]. These traditional methods struggle to detect anomalies in WSN data, which are high-dimensional and contain long-term dependencies and spatiotemporal correlations. Therefore, they often fail to fully extract the data features of WSN data. In contrast, deep learning methods can automatically extract data features. Among other deep learning methods, convolutional neural networks (CNN)[21], long short-term memory (LSTM) networks[22], and gated recurrent units (GRU)[23] have addressed the high-dimensional modeling and long-term dependency issues of time series data to some extent. However, these common deep learning methods face challenges in parallelizing their data processing strategies and often encounter gradient vanishing or explosion problems when long sequences are handled, limiting their ability to capture long-distance dependencies in time series data.

An attention mechanism capable of parallelizing data processing and directly computing the relationships between any two points in a sequence gives a transformer the ability to capture long-term dependencies. This has led to significant applications in the field of time series anomaly detection [24-26]. Nevertheless, the existing transformer-based anomaly detection methods still have shortcomings when applied to WSN data.

First, the current transformer models for WSN anomaly detection need better long-term dependency modeling capabilities for WSN sequence data. Although transformers have significantly mitigated the long-term dependency problem, they still face limitations in practical applications[10]. First, long time series contain rich temporal patterns that influence each other; thus, directly extracting long-term dependencies from long time series is not a completely reliable approach[27-30]. Second, the computational complexity of an attention mechanism increases quadratically with the sequence length, posing computational challenges when handling long sequences[27,30]. Some studies[27-30] have proposed decomposing time series to separately extract trend and seasonal components to achieve improved model performance. However, their use of moving averages for decomposition doubles the required data volume, and owing to the quadratic complexity of attention mechanisms, this increased data volume imposes a significant computational burden, limiting the ability to model long-term dependencies.

Second, most existing transformer-based anomaly detection models perform feature extraction in the time domain[31-33]. However, implementing feature extraction in the time domain results in limitations. WSN data, such as temperature and humidity data, often exhibit periodic distributions and variations (e.g., yearly, monthly, and daily cycles) and contain rich frequency components. Time-domain analysis methods struggle to reveal these intrinsic frequency structures and effectively capture the component features of time series patterns. Moreover, in the time domain, the distinction between normal and anomalous data is less pronounced than that in the frequency domain[34-36], reducing the effectiveness of anomaly detection.

Finally, the data collected by WSN sensor nodes have both temporal and graph characteristics[37]. Graph learning can serve as an important approach for capturing the spatial dependencies and intermodal dependencies in WSN data. Considering these dependencies can significantly improve the effectiveness of WSN anomaly detection[11,12]. However, the existing transformer detection methods do not comprehensively consider the spatiotemporal correlations and intermodal relationships among the data collected by sensor nodes and fail to effectively extract the correlation features among WSN time series data and detect spatiotemporal correlation anomalies. Therefore, integrating an effective graph learning method into the transformer detection approach is necessary for enhancing its anomaly detection performance.

To address the aforementioned issues, this paper proposes an anomaly detection method for WSN that integrates frequency-domain features and dynamic graph neural networks within an autoencoder reconstruction framework. First, to solve the long-term dependency problem, a new time series decomposition method is integrated into the transformer model framework. This method uses the discrete wavelet transform for multiscale time series decomposition, effectively extracting the overall trend and local details of the input time series. The downsampling operation maintains a constant data volume, reducing the computational complexity of the model and better addressing the long-term dependency modeling issue. Second, to improve the efficiency of time feature extraction in the time domain and enhance the distinction between normal and anomalous data, a frequency-domain attention mechanism (FDAM) is designed. This mechanism can more clearly capture the periodic components and frequency features of signals. Additionally, it leverages the differences between the amplitude distributions of the normal and anomalous data in the frequency domain, improving the ability of the model to identify anomalous data. Finally, to enhance the spatial modeling ability of the transformer model for WSN graph data, a multimodal fusion-based dynamic graph convolutional network (MFDGCN) is proposed, which combines an attention mechanism with a GCN. This module adaptively adjusts the network structure through the attention mechanism and uses a cross-attention mechanism to fuse features acquired from different modalities, dynamically modeling the observed spatial correlations. It learns the relationships and associations between different modalities and nodes. During anomalies, it can more accurately capture the spatial correlation changes between anomalous sensors and other sensors, thereby improving the anomaly detection performance of the model.

In summary, the main contributions of this paper are as follows:

1) The proposed WSN anomaly detection method involves the decomposition of WSN time series data into seasonal and trend components via the discrete wavelet transform. The design of the multilayer perceptron (MLP) trend encoding is intended for the seasonal component, and an FDAM mechanism is designed for the trend component, thus addressing the long-term dependency problem of WSN data derived from both seasonal and trend components. The extraction of the long-term trend of WSN data from the seasonal component ensures reliability while reducing the computational complexity of the model, thus better addressing the long-term dependency modeling issue of transformers.

2) To address the inadequacy of time-domain analysis methods in terms of extracting WSN data features, the proposed WSN anomaly detection method designs a frequency-domain attention mechanism. This mechanism fully leverages the differences between the amplitude distributions of normal and anomalous data in the frequency domain, improving the ability of the model to identify anomalous WSN data.

3) Considering the spatiotemporally correlated anomalies in WSN data, the proposed WSN anomaly detection method designs an anomaly detection module (MFDGCN) based on an attention mechanism and a GCN. This module can adaptively adjust the network structure and use a cross-attention mechanism to fuse features acquired from different modalities of WSN time series data, dynamically extracting the spatial correlation features and integrating information from different time series modalities.

## II. RELATED WORK

This section introduces the related works on solving WSN anomaly detection problems via traditional methods and deep learning methods. In the past, many researchers utilized nondeep learning methods such as digital signal processing and machine learning to address WSN anomaly detection tasks. In [17], a K-nearest neighbor (KNN) graph signal model was established on the basis of sensor location characteristics, and then a statistical test quantity was constructed based on the ratio of the smoothness of the target graph signal before and after conducting low-pass filtering. In [18], a system graph was first partitioned into subgraphs, and then the graph signals were converted to Laplacian spectral signals, processed by a low-pass filter, and restored to node-domain signals. Anomalies were detected by comparing the restored signal with the original signal. Reference[38] proposed an anomaly detection algorithm based on the K-medoid approach. Reference[19] used the k-means algorithm to divide sensor nodes into several clusters for anomaly detection. Reference[20] applied tree models to anomaly detection tasks. Reference [39] proposed an improved isolation forest algorithm using k-means and nearest-neighbor algorithms for anomaly detection purposes. In the literature [58], it is proposed to combine the ideal features of KPCA modelling with an unsupervised one-class SVM (OCSVM) scheme to discriminate between normal and abnormal measurements. These nondeep learning methods cannot extract the temporal dependencies of time series data, making it difficult to detect anomalies from the overall trend of the input time series. Additionally, traditional machine learning methods struggle with high-dimensional data, resulting in a detection accuracy bottleneck.

Deep learning methods can better extract dependencies between data and model high-dimensional data. WSN data typically has characteristics such as long temporal sequences, graph structures, and multimodal features. When designing deep learning-based neural network models for WSN data, temporal feature extraction, spatial feature extraction, and multimodal feature extraction need to be considered. In previous research, recurrent neural networks (RNN) [49] have been used to extract temporal features. However, RNN suffer from vanishing and exploding gradient problems, making it difficult to capture long-term dependencies in sequential data. To address these gradient-related issues, researchers have proposed LSTM [50-53], which introduces three gates (a forgetting gate, an input gate, and an output gate) to control the flow of information, effectively mitigating the vanishing gradient problem. Despite their advantages, LSTM models involve many parameters, leading researchers to propose GRU [54] as a simplified alternative. A GRU retains the gate mechanism of the LSTM, but reduces the number of parameters required, thereby reducing the computational time involved. Although a GRU partially solves problem of long term dependencies, it still struggles to effectively capture long range dependencies. In addition, because GRU compute each output based on the result obtained at the previous time step, they are not well suited to parallel data processing. In contrast, the transformer model, with its self-attention mechanism, can capture the dependencies between any two positions in a sequence. This ability makes it highly effective at handling long-term dependencies, leading to its prominent applications in the field of time series anomaly detection. Reference[21] designed transformer encoder layers and 1D convolution decoder layers with different scales to predict the global trend and local changes exhibited by time series. Reference[40] proposed a masked time series modeling method based on a transformer that uses a masking mechanism to reconstruct the current time step, thereby achieving reconstruction for the entire sequence. Reference[23] used a parallel transformer GRU as the information extraction module of their model to learn the long-distance correlations between timestamps and the global feature relationships of multivariate time series, using a fully connected layer for prediction and combining the prediction and reconstruction tasks to perform anomaly detection. Reference[31] generated pseudoanomalous data via a transformer to increase the amount of available anomalous data information and reconstructed it as normal data via a dual AE. Reference[32,33,41] proposed the use of a variational autoencoder (VAE) for anomaly detection. For example, [41] proposed an unsupervised anomaly detection model based on a variational transformer, using a self-attention mechanism to capture the potential correlations between sequences and providing improved position encoding and upsampling algorithms to capture multiscale temporal information. Reference[42] proposed an unsupervised learning-based variable generative adversarial transformer (VGAT) anomaly detection method, which uses a generator to reconstruct samples for anomaly detection. Reference[43] used a dual transformer structure to extract dataset association features and employed an improved adaptive multihead attention mechanism to infer trends of each dimension of multivariate time series data in parallel. Reference[44] effectively detected anomalies via an adversarial transformer structure and increased the degree of distinction between normal and anomalous data via an anomaly probability strategy. The transformer-based methods mentioned above have significantly mitigated the long-term dependency problem but still have notable limitations. In real-world data cases, long time series often contain rich temporal patterns that overlap and interact with each other, making it difficult to effectively extract temporal features. References [27-30] proposed to use a moving average approach to decompose time series into trend and seasonal components, and extract these features separately to achieve improved model performance. However, this approach significantly increases the computational complexity of the developed model. To address this issue, we propose the use of the Discrete Wavelet Transform (DWT) for time series decomposition. This approach effectively achieves the decomposition while significantly reducing the computational complexity of the model. The above methods focus on extracting temporal features in the time domain. However, through a data analysis, we have found that WSN data typically exhibit strong periodicity, which is difficult to capture effectively using time-domain analysis. Furthermore, through data observation, we found that the differences between anomalous data and normal data are much more pronounced in the frequency domain. Therefore, we propose the use of a frequency-domain attention mechanism to extract temporal features.

The spatial correlations between WSN data are also critical for WSN anomaly detection. The existing studies have often utilized GNN to extract spatial and multimodal features from WSN data. A GNN can aggregate node information on the basis of an adjacency matrix. References[11,12,24-26,45,56,57,59] utilized a GNN to capture the dependencies between variables. Reference[45] combined structural learning methods with a GNN and used the attention coefficients of adjacent sensors in the graph to predict future sensor behaviors for anomaly detection. Reference[24] proposed a new disentangled dynamic deviation transformer network (D3TN), which designs a multiscale aggregation scheme to capture the hidden dependencies between sensors and introduces a self-attention mechanism to capture the dynamic dependencies between sensors. Reference[25] used the Gumbel Softmax sampling method to learn the relationships between sensors and designed an influence propagation convolution to capture the anomalous information flows between network nodes. Furthermore, the aforementioned methods that utilize GNN to account for the spatial correlations between sensors do not take into account the fact that anomalous data can disrupt normal spatial structural relationships, rendering them unable to detect topology changes. While references [25] and [45] introduced graph structure learning, they did not make corresponding adjustments for anomalous nodes, limiting the improvement achieved in terms of spatial feature extraction. The traditional GNN also has a limited ability to capture correlations between modalities, but WSN data often exhibit strong intermodal correlations. Effectively capturing these correlations can significantly improve the resulting anomaly detection performance. However, the above methods lack effective mechanisms to enhance the ability of a GNN to capture intermodal relationships. To address these shortcomings, we propose a graph learning approach that uses attention mechanisms to adaptively adjust the network structure and cross-attention mechanisms to integrate features derived from different modalities.

In summary, the following work will be carried out to address the deficiencies of the existing WSN anomaly detection methods. By introducing new time series decomposition techniques and an FDAM, the ability to capture long-term trends and periodic features is improved, enhancing the accuracy of anomaly detection. Additionally, by combining an attention mechanism with a GCN, an MFDGCN is designed, which can adaptively and fully extract the spatial correlation features of WSN, effectively integrating the information of different modalities and nodes to achieve improved anomaly detection performance.

## III. PROBLEM STATEMENT

Following a similar approach to that in reference [37], this paper represents multinode, multimodal WSN time series data as spatiotemporal graph data, which can be expressed as a dynamic attributed graph $\mathbf{G} = \{\mathbf{G}_1, \mathbf{G}_2, \ldots, \mathbf{G}_T\} \in \mathbb{R}^{N \times M \times T}$. Here, $\mathbf{G}_t = (\mathbf{X}_t, \mathbf{A}_t)$ represents a snapshot of the attributed graph $\mathbf{G}$ at time $t$, where $\mathbf{X}_t \in \mathbb{R}^{N \times M}$ and $\mathbf{A}_t \in \mathbb{R}^{N \times N}$ are the attribute matrix and adjacency matrix of $\mathbf{G}_t$, respectively. $N$ represents the number of nodes contained in the sensor network, $M$ represents the number of time series variables (or modalities) measured by the sensor nodes, and $T$ represents the process of sampling over $T$ time points. The adjacency matrix $\mathbf{A}_t$ reflects the spatial topology of the network graph, where the element $\mathbf{A}_t(i,j)$ indicates the correlation between nodes $i$ and $j$. The $i$-th row of the attribute matrix $\mathbf{X}_t$, denoted as $\mathbf{X}_t^i \in \mathbb{R}^M$, represents the $M$-dimensional modal time series data of sensor node $i$ at time $t$.

The WSN anomaly detection task in this paper involves analyzing the input WSN attribute graph data to determine which node at which time point has produced anomalous data deviating from the normal distribution. A successful WSN anomaly detection strategy must fully leverage the correlations within the data, including the spatial similarity differences caused by the spatial positions of the sensor nodes, the correlations among different modalities within a sensor node, and the temporal correlations across different time points. This paper aims to comprehensively exploit these correlations and model the anomalous WSN data stream detection problem as a classification task. We take $W$ consecutive time windows of sensor network observations as inputs: $\mathbf{G} = \{\mathbf{G}_1, \mathbf{G}_2, \ldots, \mathbf{G}_W\}$. Designing an appropriate anomaly detection model involves determining a suitable neural network structure and corresponding weight parameters $\theta$, with the mapping function denoted as $f$. Through the function $f$, the input WSN data $\mathbf{G}$ are used to determine whether the reading of modality $j$ at node $i$ at time $t$ is anomalous; this is expressed as:

$$\mathbf{Y} = f(\mathbf{G} \mid \theta) \in \{0,1\}^{N \times M \times W} \qquad (1)$$

The reading of modality $j$ at node $i$ at time $t$ can be denoted as $y_t(i,j)$, where $y_t(i,j) = 1$ indicates an anomaly in the WSN data at node $i$ and modality $j$ at time $t$, and $y_t(i,j) = 0$ indicates that no anomaly is present.

## IV. METHODOLOGY

The WSN anomaly detection model designed in this paper, as shown in Figure 1, adopts an autoencoder network as its basic framework. It consists of four main components: a time series decomposition module, a trend encoder, a seasonal encoder, and a decoder.

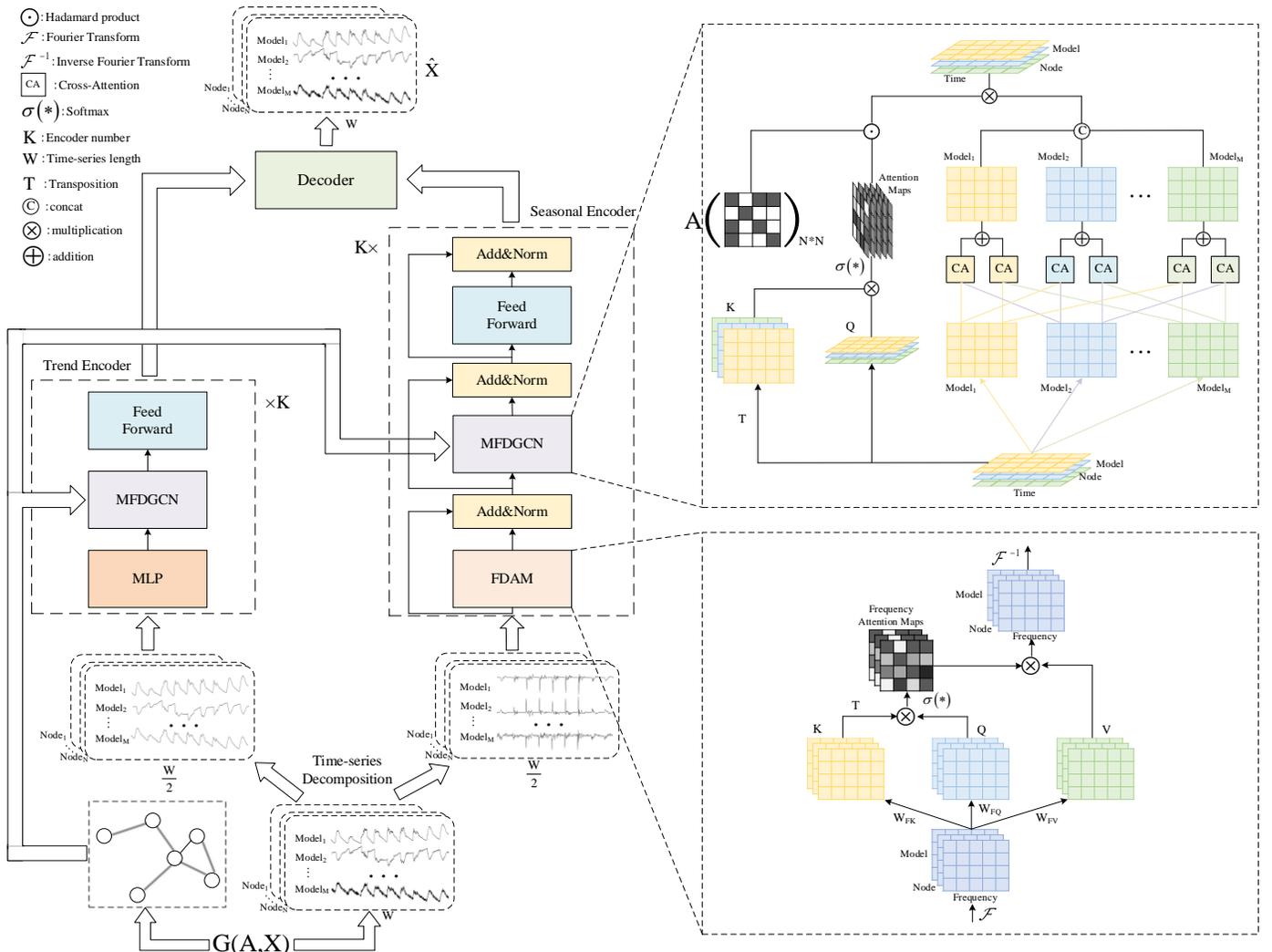

Fig. 1. Model Framework Diagram

The time series decomposition module uses the discrete wavelet transform to decompose the time series data of each modality at each WSN node into trend and seasonal components. Reference [28] theoretically and empirically demonstrated that the effectiveness of attention mechanisms varies with different temporal patterns. For seasonal data, an attention mechanism is more effective, whereas for trend data, as attention mechanisms essentially interpolate context, they cannot effectively extrapolate linear trends and thus perform poorly. Therefore, the trend encoder uses an MLP to better predict and capture the trend signal features, whereas the seasonal encoder employs an attention mechanism for feature extraction purposes.

To leverage the strengths of a transformer while endowing it with graph learning capabilities, this paper integrates the attention mechanism with GNN and designs a multimodal fusion-based dynamic graph convolution module. The decoder employs an inverse discrete wavelet transform to reconstruct the encoded trend and seasonal components back into their original time series form, and then linear mapping is applied to obtain the reconstructed output.

The design and analysis of each module are as follows:

### A. Time Series Decomposition Module

The wavelet transform is a commonly used time–frequency-domain analysis method. It can decompose a time series into different subseries on the basis of various frequency components, retaining both the frequency and time information of the original series. This allows for decomposing the series from both temporal and frequency perspectives, thus capturing more information contained in the given time series data. The wavelet transform can also perform multiscale refinement, offering high frequency resolutions and low temporal resolutions at low frequencies and high temporal resolutions and low frequency resolutions at high frequencies. This capability allows it to focus on any detail of the input signal and makes it particularly sensitive to abrupt changes in time series data; thus, it is advantageous for detecting anomalies in WSN data.

Wavelet transforms can be categorized into the continuous wavelet transform (CWT) and the discrete wavelet transform (DWT). A wavelet transform involves two variables: a scale factor $a$ and a translation factor $b$. The scale factor $a$ controls the dilation of the wavelet function $\psi$, and the translation factor $b$ controls the translation of the wavelet function $\psi$. The DWT discretizes both $a$ and $b$. In binary discretization, the values of $a$ and $b$ are simplified to powers of 2. The discrete wavelet function at this point is as follows:

$$\psi_{j,k}(t) = 2^{-\frac{j}{2}} \psi(2^{-j}t - k) \quad (2)$$

where $\forall j, k \in Z$.

The complete formula for the DWT of an input signal $x(t)$ is as follows:

$$DWT_\psi(j,k) = 2^{\frac{j}{2}} \int_{-\infty}^{+\infty} x(t) \psi(2^j t - k) dt \quad (3)$$

This paper uses the Mallat algorithm (4)(5) to achieve multilevel decomposition with the DWT.

$$A_{j+1}(n) = H(n) * A_j(n) = \sum_k H(k) A_j(2n-k) \quad (4)$$

$$D_{j+1}(n) = G(n) * A_j(n) = \sum_k G(k) A_j(2n-k) \quad (5)$$

where $H$ is a low-pass filter, $G$ is a high-pass filter, $A_{j+1}$ represents the approximation component of the $j+1$-th decomposition level, and $D_{j+1}$ represents the detail component of the $j+1$-th decomposition level.

The reconstruction formula that uses the Mallat algorithm to reconstruct discrete sequences back into their original forms is as follows:

$$\begin{aligned} A_j(n) &= h(n) * A_{j+1}(n) + g(n) * D_{j+1}(n) \\ &= \sum_k h(n-2k) A_{j+1}(k) \\ &\quad + \sum_k g(n-2k) D_{j+1}(k) \end{aligned} \quad (6)$$

where $h(n)$ and $g(n)$ represent the tap coefficient sequences of the low-pass and high-pass filters, respectively, corresponding to the selected wavelet function.

The block diagram of the Mallat algorithm is shown in Figure 2, where $h$ represents low-pass filtering, $g$ represents high-pass filtering, and $\downarrow 2$ indicates downsampling by a factor of 2. The two complementary low-pass and high-pass filters can separate the low-frequency and high-frequency components of the input time series, obtaining an approximation component $A$ and a detail component $D$ after downsampling by a factor of 2. This strategy retains all the information acquired from the original time series without increasing the data volume. During reconstruction, the approximation and detail components are recombined via the tap coefficients to restore the original signal.

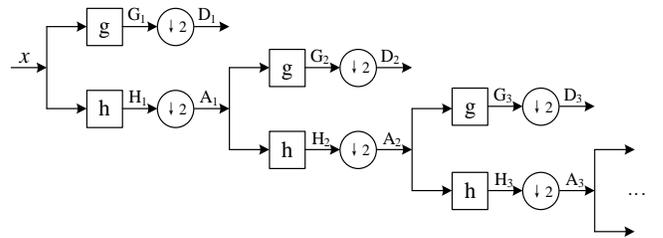

Fig. 2. Block Diagram of the Mallat Algorithm

On the basis of the above analysis, this paper adopts the first-order DWT for trend and seasonal decomposition in the designed model framework (as shown in Figure 1). The trend component of a time series represents its overall development trend, and the general process of the time series changes and belongs to the low-frequency component. The seasonal component represents the fluctuations in time series details, which are relatively small and rapidly changing and belong to the high-frequency component. Therefore, after performing the first-order DWT on the original time series, the approximation component is used to represent the trend component, and the detail component is used to represent the seasonal component. The trend and seasonal components are then input into the trend encoder and seasonal encoder, respectively, for feature extraction purposes. The time series decomposition module can be represented as:

$$\mathbf{X}_{tre}, \mathbf{X}_{sea} = \mathrm{DWT}(\mathbf{X}) \quad (7)$$

where $\mathbf{X} \in \mathbb{R}^{N \times M \times W}$ represents the input time series, $\mathbf{X}_{tre} \in \mathbb{R}^{N \times M \times \frac{W}{2}}$ represents the trend component, $\mathbf{X}_{sea} \in \mathbb{R}^{N \times M \times \frac{W}{2}}$ represents the seasonal component, and $\mathrm{DWT}(*)$ represents the discrete wavelet transform.

### B. Trend Encoder Module

The trend encoder in the designed model framework (as shown in Figure 1) mainly comprises an MLP and a multimodal fusion-based dynamic graph convolution module. The MLP consists of three layers: an input layer, a hidden layer, and an output layer. The input layer receives input data, the hidden layer learns feature representations, and the output layer produces the final output. Each layer is fully connected to the next layer, and each neuron in the hidden layers has a nonlinear activation function for introducing nonlinear mappings. Since each layer of the MLP is fully connected to the next layer, the output obtained for each point comprehensively considers all the input points, providing a good level of perception regarding the overall trend of the time series. Therefore, the trend encoder using the MLP can effectively model the time dependencies of the given data.

The designed multimodal fusion-based dynamic graph convolution module models the spatial dependencies of WSN data. At each sampling moment, the sensor data and their topological structure form a graph. This implies a certain degree of correlation among the sensors, but classic transformers cannot directly learn graph-structured data to capture the spatial information of WSN data. Therefore, this paper combines an attention mechanism with a GCN to design an MFDGCN module for modeling spatial dependencies. This module uses node attributes to compute node associations with a self-attention mechanism, adaptively adjusting the network topology for anomaly detection tasks. To extract the mutual dependencies among different modalities within a node, the MFDGCN uses cross-attention mechanisms for multimodal information fusion.

In a WSN, the spatial correlations of data are crucial for anomaly detection. Anomalies can disrupt the spatial relationships contained in WSN data, but traditional GCN fail to capture these spatial dependency changes caused by anomalies. GCN aggregate features from neighboring nodes to extract information, which helps to extract local features for normal data but may obscure the key differences between anomalous and normal nodes when anomalies occur, reducing the sensitivity of the model to anomalies. Thus, adjusting the network topology accordingly is necessary. This paper uses a self-attention mechanism to dynamically compute the association strengths between nodes and then computes the Hadamard product of the spatial correlation weight matrix and the adjacency matrix to adjust the network topology. At time $t$, the node representation $\mathbf{Z}_t \in \mathbb{R}^{N \times M}$ and the spatial correlation weight matrix $\mathbf{S}_t$ are represented as follows:

$$\mathbf{S}_t = \text{Softmax}\left(\frac{\mathbf{Z}_t \mathbf{Z}_t^{\text{T}}}{\sqrt{d_M}}\right) \in \mathbb{R}^{N \times N} \quad (8)$$

The spatial correlation weight matrix $\mathbf{S}_t$ and the adjacency matrix $\mathbf{A}$ are then elementwise Hadamard multiplied to adjust the adjacency matrix:

$$\tilde{\mathbf{A}} = \mathbf{S}_t \odot \mathbf{A} \quad (9)$$

where $\odot$ represents the Hadamard product and $\tilde{\mathbf{A}}$ is the adjusted adjacency matrix.

The cross-attention mechanism calculates the attention between two different sequences to obtain the dependencies between them. This paper uses a cross-attention mechanism to fuse different modal sequences, mapping one modal sequence to a query (Q) and another different modal sequence to a key (K) and value (V). By computing attention values and summing them, this process fuses information from other modalities. Given an input attribute matrix $\mathbf{X} \in \mathbb{R}^{N \times M \times T}$, the matrix can be modally expanded as $\mathbf{X} = \{\mathbf{X}_1, \mathbf{X}_2, ..., \mathbf{X}_M\} \in \mathbb{R}^{N \times T}$. The fusion of modality $i$ with other modalities can be represented as:

$$\text{fusion}_i = \sum_{j=1, j \neq i}^{M} \text{Softmax}\left(\frac{\mathbf{X}_j \mathbf{W}_Q (\mathbf{X}_i \mathbf{W}_K)^{\text{T}}}{\sqrt{d_M}}\right) \mathbf{X}_i \mathbf{W}_V \quad (10)$$

The modal fusion process is represented as:

$$\begin{aligned}\text{ModalFusion}(\mathbf{X}) = \\ \text{Concat}(\text{fusion}_1, ..., \text{fusion}_M)\end{aligned} \quad (11)$$

The propagation between layers in the multimodal fusion-based dynamic graph convolution module is represented as follows:

$$\mathbf{H}^{(l+1)} = \sigma\left(\tilde{\mathbf{A}} \text{ModalFusion}(\mathbf{H}^{(l)}) \mathbf{W}^{(l)}\right) \quad (12)$$

where $\sigma(*)$ represents the activation function and $\mathbf{H}^{(l)}$ represents the feature representation of the $l$-th layer.

The trend encoder designed in this way combines the attention mechanism with a GCN, adaptively adjusting the network topology and fusing multimodal information. This significantly improves the accuracy of anomaly detection and the adaptability of the model to dynamic changes, enhancing its sensitivity to anomalous data.

### C. Seasonal Encoder Module

The seasonal encoder comprises an FDAM and an MFDGCN module. The structure of the MFDGCN is the same as that described in Section 4.1, where it models the spatial dependencies of WSN data. Therefore, this section primarily introduces the designed FDAM.

Through data analysis, it has been found that for anomaly detection tasks involving WSN data, frequency-domain feature extraction methods are more efficient than traditional time-domain feature extraction methods. This is because WSN data typically exhibit significant periodicity and contain distinct frequency components. In the frequency domain, owing to their periodicity, normal data exhibit concentrated energy at specific frequency components, occupying a dominant position. In contrast, owing to their nonperiodicity, anomalous data have dispersed energy across the frequency spectrum, making distinguishing between normal and anomalous data through frequency-domain analysis easier.

For example, consider the cases of point anomalies, contextual anomalies, and collective anomalies. As shown in Figure 3, the original time series includes anomalous data, which can be considered the sum of normal data and anomalies. After performing a Fourier transform on these three time series, the corresponding frequency spectra are obtained, as shown in Figure 4. Figure 4 clearly shows that the frequency components of the normal data are mainly concentrated in the low-frequency range, whereas the frequency components of the anomalous data are more dispersed across the entire spectrum. Since the Fourier transform is linear, the spectrum of the original time series is the sum of the spectra of the normal and anomalous series. The similarity between the original time series spectrum and the normal time series spectrum indicates that the frequency components of the normal data are more concentrated and occupy most of the spectrum, which is consistent with our analysis.

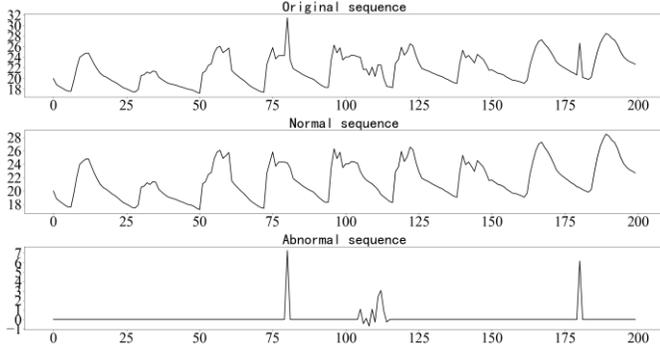

Fig. 3. Time Series Plots

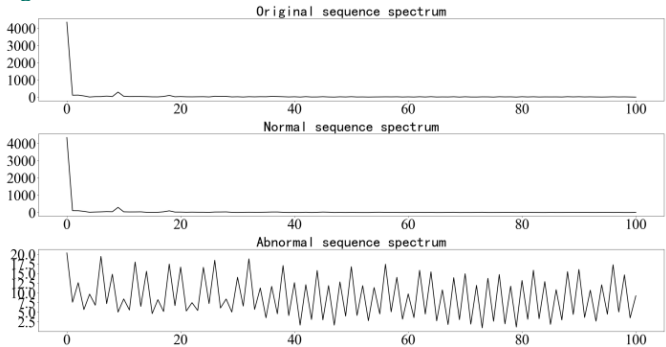

Fig. 4. Frequency Spectrum Plots

Given that normal and anomalous data exhibit different characteristics in the frequency domain, it is feasible to extract sequence features for anomaly detection purposes from the frequency domain. Therefore, this paper designs an FDAM to extract features from the frequency domain, capturing the periodic information of the sequence to detect anomalies.

First, it is necessary to transfer the original sequence from the time domain to the frequency domain. The Fourier transform is a common method for converting a time series from a time-domain series to a frequency-domain series. It decomposes the time input series into a series of trigonometric functions, which represent the amplitude and phase of the time series at different frequencies. This paper uses the discrete Fourier transform (DFT) (13) to transfer the time series from the time domain to the frequency domain.

$$X(k) = \mathcal{F}(x(n)) = \sum_{n=0}^{N-1} x(n) e^{-i\omega kn} \quad (13)$$

It can be concisely represented as:

$$X_F = \mathcal{F}(x) \quad (14)$$

where $\mathcal{F}$ represents the DFT, $N$ is the length of the time series, $e$ is the base of the natural logarithm, $i$ is the imaginary unit, and $\omega$ represents the angular frequency.

Next, three learnable parameters in the frequency domain, $\mathbf{W}_{FQ}$, $\mathbf{W}_{FK}$, and $\mathbf{W}_{FV}$, are utilized to map $X$ in the frequency domain to queries $Q$, keys $K$, and values $V$, respectively, to enable feature extraction in the frequency domain.

$$\mathbf{Q} = \mathbf{X}_F \mathbf{W}_{FQ} \quad (15)$$

$$\mathbf{K} = \mathbf{X}_F \mathbf{W}_{FK} \quad (16)$$

$$\mathbf{V} = \mathbf{X}_F \mathbf{W}_{FV} \quad (17)$$

In anomaly detection tasks, it is usually desirable for a model to learn and reconstruct normal data while suppressing or ignoring anomalous data. This paper finds that using a self-attention mechanism in the frequency domain can achieve this effect. In the frequency spectrum, normal data typically have larger and more concentrated amplitudes, whereas anomalous data have smaller and more dispersed amplitudes. Therefore, the self-attention mechanism can easily focus on the frequency components of normal data, assigning them higher attention weights. When normalizing attention weights, the Softmax function is typically used. The Softmax function, with its exponential term, has a "polarizing" effect, increasing the gap between larger and smaller values and giving larger weights to higher values and smaller weights to lower values. Hence, using an attention mechanism in the frequency domain can better focus attention on the frequency components of the normal data while suppressing the frequency components of the anomalous data. This effectively captures the features of the normal data and reduces the impact of the anomalous data.

Finally, the inverse DFT is used to transfer the sequence from the frequency domain back to the time domain. Its formula is as follows:

$$x(n) = \mathcal{F}^{-1}(X(k)) = \frac{1}{N} \sum_{k=0}^{N-1} X(k) e^{i\omega kn} \quad (18)$$

where $\mathcal{F}^{-1}$ represents the inverse DFT.

Thus, the frequency-domain attention mechanism can be expressed as:

$$\text{FDAM}(\mathbf{Q}, \mathbf{K}, \mathbf{V}) = \mathcal{F}^{-1}\left(\text{Softmax}\left(\frac{\mathbf{Q}\mathbf{K}^{\text{T}}}{\sqrt{d_k}}\right)\mathbf{V}\right) \quad (19)$$

Through this design, the periodicity and frequency characteristics of WSN data signals can be explored in detail in the frequency domain, significantly improving the efficiency and accuracy of the feature extraction process. Additionally, this approach cleverly leverages the significant difference between the amplitude distributions of the normal and anomalous data, enhancing the sensitivity of the model to anomalous data.

### D. Decoder Module

The decoder, as designed in Figure 1, consists of an inverse discrete wavelet transform (IDWT) and a linear layer. The IDWT is used to reconstruct the original time series data from the encoded trend and seasonal components via equation (6). The final output is obtained through a linear mapping of the reconstructed components. The decoder can be represented as follows:

$$\hat{\mathbf{X}} = \text{Linear}(\text{IDWT}(\mathbf{Z}_{tre}, \mathbf{Z}_{sea})) \quad (20)$$

where $\mathbf{Z}_{tre}$ denotes the encoded trend component, $\mathbf{Z}_{sea}$ denotes the encoded seasonal component, $\text{IDWT}(*)$ denotes the inverse discrete wavelet transform, $\text{Linear}(*)$ denotes the linear layer, and $\hat{\mathbf{X}}$ represents the reconstructed time series.

### E. Anomaly Score

Since the proposed method is a reconstruction-based anomaly detection approach, its goal is to learn the characteristics of normal data by minimizing the mean squared error (MSE) between the reconstructed output $\hat{\mathbf{X}}$ and the actual observed input data $\mathbf{X}$. The loss function is defined as shown below:

$$\mathcal{L}_{mse} = \frac{1}{WNM}\sum_{t=1}^{W}\sum_{i=1}^{N}\sum_{j=1}^{M}\left(x_t(i,j)-\hat{x}_t(i,j)\right)^2 \quad (21)$$

During testing, anomalous data are input into the model, which reconstructs these data points. Since the distribution of anomalous data differs from that of normal data, the reconstruction error induced for anomalous data is relatively large. Therefore, the reconstruction error, computed as the squared difference between the input value and the reconstructed value, is used as the anomaly score. The anomaly score for the $j$-th mode at node $i$ at time $t$ is represented as:

$$\text{score}_t(i,j) = \left(x_t(i,j)-\hat{x}_t(i,j)\right)^2 \quad (22)$$

The anomaly score is compared with a threshold $\tau$. If the anomaly score exceeds this threshold, the data point is labeled as an anomaly (its label is set to 1); otherwise, it is labeled as normal (its label is set to 0). Usually, the threshold $\tau$ setting was determined using the training set, which consists solely of normal data. Consequently, we used the highest anomaly score from the training set as the detection threshold.

$$y_t(i,j) = \begin{cases} 1, & \text{score}_t(i,j) > \tau \\ 0, & \text{score}_t(i,j) \le \tau \end{cases} \quad (23)$$

## V. Experiments

### A. Experimental Dataset and Environment

The Intel Berkeley Research Lab dataset (IBRL) [48] was utilized for experimental validation. The IBRL dataset, which was collected by Intel Berkeley Research Lab, consists of data from 54 sensors distributed across the laboratory, as illustrated in Figure 5. This dataset includes readings of humidity, temperature, light, and voltage values recorded every 31 seconds over more than a month. Before using this dataset, we carried out extensive analysis and preprocessing work. First, we found that the data from node 5 and node 15 had many missing values, so we excluded these two nodes and considered the data from the remaining 52 sensor nodes. Second, the quality of the illumination data was very poor, so we only used the humidity, temperature, and voltage modalities. Finally, we identified unreasonable data points, such as those with indoor temperatures exceeding 120°C or negative humidity values. These unreasonable data points were removed to obtain a clean and valid dataset.

Before analyzing the processed data, it was necessary to normalise the data to remove the effects of differences in units of measurement and range of values between the modalities. This required the raw data to be standardized so that all the modalities were on the same scale. In this study, we used Z score normalization to standardize the input tensors:

$$\mathbf{X}_{ij} = \frac{\mathbf{X}_{ij}-\mu(\mathbf{X}_{ij})}{\sigma(\mathbf{X}_{ij})} \quad (24)$$

Here, $\mathbf{X} \in \mathbb{R}^{N \times M \times W}$ represents the input tensor, and $\mathbf{X}_{ij}$ represents the time series data recorded by the $j$-th sensor at the $i$-th node. $\mu(\mathbf{X}_{ij})$ is the mean, and $\sigma(\mathbf{X}_{ij})$ is the standard deviation of $\mathbf{X}_{ij}$. After performing Z score preprocessing, the modal time series data followed a standard normal distribution with a mean of 0 and a standard deviation of 1.

The experimental setup was configured with the following hardware and software specifications: Intel(R) Xeon(R) Gold 5218 CPU @ 2.30 GHz, NVIDIA GeForce RTX 3090 GPU, Ubuntu 18.04.2 LTS operating system, Python 3.6.9, PyTorch 2.0.0, and CUDA Version 11.7. The code scripts are provided on an open-source platform.[1]

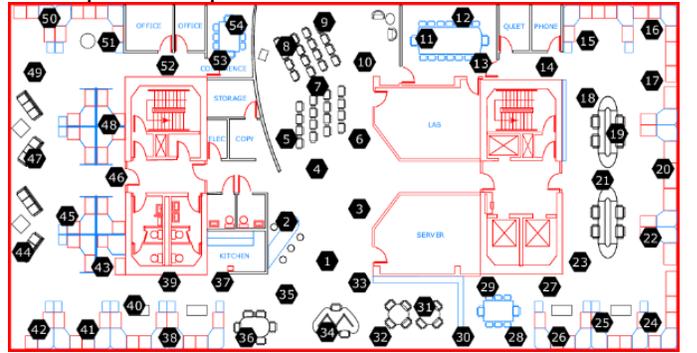

Fig. 5. Spatial distribution of sensors in the IBRL dataset

### B. Experimental Model Parameters and Evaluation Metrics

Appropriate values for the learning rate and number of training epochs can lead to rapid convergence and optimal training performance. The sliding window size must balance the long-term dependencies of the input time series data with the computational complexity of the model. Longer time series inputs provide more information to the model, thereby improving its anomaly detection accuracy. However, since the designed model uses an attention mechanism whose computational complexity increases quadratically with the sequence length, excessively long sliding windows lead to prohibitively high computational costs. Considering these factors, we set the sliding window size W to 300. For the sliding step size, the experimental results showed that the reconstruction performance achieved for the data in the second half of the sliding window was better than that achieved for the data in the first half, probably due to the influence of the long-term dependencies contained in the time series data. Since the model had to first learn to extract time series features before it could effectively reconstruct the data, we decided to use only the last 100 time points of the sliding window for anomaly detection. Thus, the sliding step size L was set to 100. In addition, we conducted several rounds of experiments for the learning rate r $\in [0.0001, 0.0002, \cdots, 0.01]$ and training epochs epoch ∈ [100,150,200,300,400]. Based on the above experimental trials and analysis, in the experiments, the Adam optimizer was employed, and after multiple trials, the learning rate was determined to be r=0.001, with 200 training epochs. The sliding window W was set to 300, and the sliding step size L was set to 100.

---

[1] The source code is publicly available at https://github.com/JZBgit/wsnad .

The objective of the task is to determine the presence of anomalies in the data. For such a binary classification task, the following four scenarios are usually considered: true positives (TP), true negatives (TN), false positives (FP), and false negatives (FN). In anomaly detection, anomalous data are considered positive P, and normal data are considered negative N. Therefore, TP denotes the number of samples correctly predicted as anomalies, TN denotes the number of samples correctly predicted as normal, FP denotes the number of normal samples incorrectly predicted as anomalies, and FN denotes the number of anomalies incorrectly predicted as normal. The performance metrics for classification are summarized in the confusion matrix shown in Table 1.

TABLE I
Classification Confusion Matrix

|  | Predicted Anomaly | Predicted Normal |
|---|---|---|
| True Anomaly | TP | FN |
| True Normal | FP | TN |

Precision measures the proportion of correctly predicted positive samples (TP) among all samples predicted as positive (TP + FP). The formula for precision is:

$$\text{Precision} = \frac{\text{TP}}{\text{TP} + \text{FP}} \tag{25}$$

Recall measures the proportion of correctly predicted positive samples (TP) among all actual positive samples (TP + FN). The formula for the recall is:

$$\text{Recall} = \frac{\text{TP}}{\text{TP} + \text{FN}} \tag{26}$$

In model evaluation, it is desirable for the model to have high recall (i.e., to detect as many anomalies as possible) and high precision (i.e., for the detected anomalies to be true anomalies). Therefore, both precision and recall are important performance metrics. Since anomaly detection samples are usually imbalanced, considering both metrics is crucial. The F1 score is a metric that combines precision and recall, balancing the trade-off between them, and is thus commonly used in anomaly detection tasks. The formula for the F1 score is:

$$\text{F1} = 2 \times \frac{\text{Precision} \times \text{Recall}}{\text{Precision} + \text{Recall}} \tag{27}$$

### C. Comparative Experiments

To verify the effectiveness of the proposed method, it was compared with the following four existing methods:

(1) CNN-LSTM[46]: This method combines one-dimensional convolutional neural networks (CNN) and long short-term memory (LSTM) networks to perform multistep forecasting. It achieves this by iteratively utilizing the outputs from preceding stages in conjunction with the data recorded at the current time step. Anomalies are detected through multistep prediction.

(2) MTAD-GAT[47]: This method consists of a one-dimensional CNN, graph attention networks (GAT), and gated recurrent units (GRU), which combine reconstruction and prediction methods for multivariate time series anomaly detection. The method preprocesses each time series using one-dimensional convolution and then designs two GAT to model the temporal and feature dimensions of multivariate time series. The original input is combined with the output from the GAT and then fed into the GRU for reconstruction and prediction.

(3) GAT-GRU[11]: This method utilizes a GAT to extract features from temporal, modality, and spatial dimensions and uses a GRU for long-term dependency modeling, employing reconstruction to detect anomalies. Specifically, GAT are initially utilized to extract features from temporal and modality dimensions for each node and then use a GRU to model long-term dependencies and reduce dimensionality. The spatial features of the reduced data are extracted via the GAT, and the data are reconstructed. Large reconstruction errors indicate anomalies.

(4) GLSL[12]: This method divides WSN data into multiple graph datasets based on the modality dimension. It uses GAT to extract temporal and spatial features of each modality's graph data and then uses a GRU to extract long-term dependency features. The method combines reconstruction and prediction methods to detect anomalies.

A comparison among these methods highlights the strengths and weaknesses of the proposed method in terms of accuracy and performance in detecting anomalies in WSN data.

The experimental results of the proposed method, compared with those of the four baseline methods, are shown in Table II. Here, Pre represents precision, Rec represents recall, and F1 represents the F1 score. Par denotes the number of parameters, FLOPs represents the computational cost in terms of floating-point operations, and Time is to the inference time for a time window W = 300. AUC denotes the area under the ROC curve. AUC represents the area under the ROC curve.

As illustrated in Table II, the CNN-LSTM method achieved the lowest F1 score. This is because this method uses a CNN to extract relationships between modalities in a manner similar to image processing, which does not consider complex intermodal associations and fails to handle graph data with intricate internode relationships. Therefore, it produced the worst performance among the compared methods. In addition, as this method uses LSTM for temporal feature extraction, it suffers from problems such as many parameters and high computational complexity. Therefore, it also performed the worst in terms of the Par, FLOPs, and time metrics compared to the other methods.

The MTAD-GAT method emphasizes anomaly detection at the individual node level while disregarding the spatial correlation features among multiple nodes. This method necessitates the construction of multiple models to train and learn the features of each node, reducing training efficiency. Additionally, since MTAD-GAT does not account for internode associations, it does not achieve optimal results. In addition, since this method uses a GRU for temporal feature extraction, which has fewer parameters than does LSTM, it significantly outperformed the CNN-LSTM method in terms of the Par, FLOPs, and time metrics.

TABLE II
Comparative Experimental Results

| Method | Pre (%) | Rec (%) | F1 (%) | AUC | Par (M) | FLOPs (M) | Time (ms) |
|---|---|---|---|---|---|---|---|
| CNN-LSTM | 79.5 | 70.0 | 74.5 | 0.70 | 27.9 | 14,699.7 | 817.5 |
| MTAD-GAT | 77.5 | 87.0 | 82.0 | 0.81 | 1.1 | 749.2 | 848.8 |
| GAT-GRU | 93.3 | 87.5 | 90.3 | 0.84 | 36.5 | 14,445.6 | 13,017.0 |
| GLSL | 94.5 | 87.0 | 90.6 | 0.93 | **0.6** | 3,075.1 | 2,909.3 |
| **ours** | **94.7** | **92.3** | **93.5** | **0.97** | 0.9 | **148.6** | **271.0** |

Conversely, both GAT-GRU and GLSL demonstrated superior performance in extracting data features from temporal, modality, and spatial dimensions using different approaches. GAT-GRU first extracts temporal and modality features from each node and then concatenates all nodes to learn spatial features. GLSL decomposes data into multiple graphs by modality, extracts temporal and spatial features first, and then concatenates them to learn intermodal features. GLSL shows a distinct advantage in scenarios involving large-scale data, as GAT-GRU requires a feature extractor for each node. As shown in Table 2, although the GAT-GRU method achieved significant improvements in F1 scores over the previous two methods, it suffered from large parameter counts, high computational complexity, and long inference times. This is because GAT-GRU designs independent GAT and GRU networks for each node to extract information, resulting in very high parameter counts and computational complexity levels. As a result, GAT-GRU had the longest inference time of all the methods compared. The GLSL method divides WSN data into multiple graph datasets according to their modalities and uses GAT and GRU networks to extract information. This approach significantly reduces the number of parameters and lowers the computational complexity levels. However, GLSL still faces challenges such as high computational complexity and long inference times.

The proposed model outperformed the baseline methods in terms of precision, recall, and F1 score. It addresses the limitations identified in the above methods by incorporating time series decomposition, frequency domain feature extraction, and dynamic spatial modeling. The ablation study will demonstrate the impact of these factors on the model's performance and effectiveness. In addition, the proposed model uses an MLP and attention mechanisms to extract temporal features. This approach allows the model to capture information from different time steps and enables parallel processing of time series data during the training and inference phases. As a result, the model achieves exceptional computational efficiency and parallel processing capabilities. As shown in Table II, although the number of parameters of the proposed model is not the smallest, it is smaller than that of all methods except GLSL. Furthermore, the proposed model outperforms all baseline methods in terms of the computational complexity and inference time measures.

To assess the ability of each model to discriminate between abnormal and normal samples at different thresholds, we also tested the AUC metric for each model. AUC (Area Under the Curve) is the area under the ROC curve, where the ROC curve plots the False Positive Rate (FPR) on the x-axis and the True Positive Rate (TPR) on the y-axis. As the threshold changes from high to low, we obtain a series of (FPR, TPR) points that form the ROC curve. The AUC summarizes the TPR and FPR metrics and reflects the overall performance of the model at different thresholds. The higher the AUC, the better the model's ability to discriminate between abnormal and normal samples. After performing statistical tests, the AUC values of each model are shown in Table 2. From Table 2, it can be seen that the AUC value of the model we designed is the highest, indicating that our model has the strongest ability to performing between abnormal and normal samples.

### D. Ablation Experiments

To examine the effectiveness and performance of each component in the proposed model, we conducted ablation experiments. By systematically removing each component and observing the changes in various metrics, we were able to determine the impact of each component. The results of the ablation experiments are presented in Table III. Here, Pre represents precision, Rec represents recall, F1 represents the F1 score, Par represents the number of parameters (in millions), and FLOPs represent the computational cost in terms of floating-point operations (in millions).

TABLE III
Ablation Experimental Results

| Scheme | DWT | FDAM | MFDGCN | Pre (%) | Rec (%) | F1 (%) | Par (M) | FLOPs (M) |
|---|---|---|---|---|---|---|---|---|
| **1** | √ | √ | √ | **94.7** | **92.3** | **93.5** | **0.9** | **148.6** |
| 2 | × | √ | √ | 94.5 | 92.2 | 93.4 | 1.8 | 324.9 |
| 3 | × | × | √ | 89.1 | 84.8 | 86.8 | 1.8 | 324.9 |
| 4 | × | × | × | 79.0 | 84.7 | 81.6 | 1.7 | 240.6 |

Scheme 1 is the complete model proposed in this paper. This model employs DWT for time series trend and seasonal decomposition and then inputs them separately into the trend encoder and seasonal encoder, with the seasonal encoder using

FDAM for feature extraction. Both encoders use the MFDGCN for spatial feature extraction. Scheme 1 achieved the best performance in terms of precision, recall, and F1 score.

In Scheme 2, we replaced DWT time series decomposition with the commonly used moving average method to investigate the performance improvement brought by using the DWT. Our analysis indicates that using DWT reduces the data volume by half compared with the moving average method. With reduced data volume, the required computation and parameter size for the model are also reduced. Scheme 2 demonstrates that both the DWT and moving average methods perform effectively, with the DWT performing slightly better. Specifically, the DWT method has a computational cost of 148.6 M FLOPs and a parameter size of 0.9 M, whereas the moving average method has 324.9 M FLOPs and 1.8 M parameters. This suggests that the DWT method significantly reduces the model's computational and spatial complexity.

Scheme 3 replaces the frequency domain attention mechanism used in Scheme 2 with a time-domain signal attention mechanism to evaluate the effectiveness of the frequency domain attention mechanism. Compared with Scheme 2, Scheme 3 shows that the frequency domain attention mechanism improves precision by 5.4%, recall by 7.4%, and the F1 score by 6.6%. These improvements demonstrate the unique advantages of the frequency domain attention mechanism in enhancing the model's anomaly detection capability.

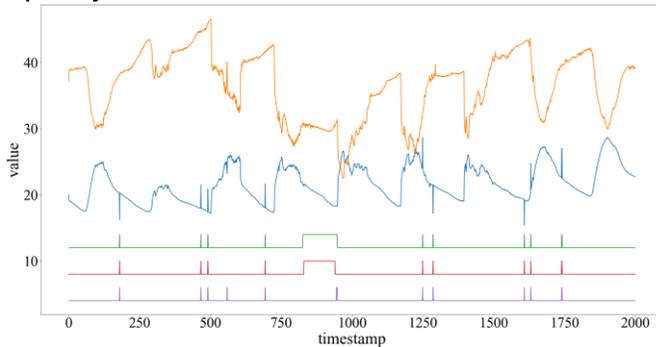

Fig. 6. Correlation Anomaly Detection Analysis

Scheme 4 replaces the MFDGCN with a standard GCN in Scheme 3 to assess the impact of dynamic spatial modeling and multimodal fusion on anomaly detection. Scheme 4 retains the GCN for static graph learning. The results indicate that the dynamic spatial learning method significantly improves the precision by 10.1% and the F1 score by 5.2%, highlighting the importance of dynamic spatial structure learning for effective anomaly detection. To further investigate the ability of the MFDGCN to extract correlation information and its effectiveness in detecting correlation anomalies, we analyzed the model's actual anomaly detection performance. Figure 6 shows the temperature (blue line) and humidity (orange line) data of node 1 in the IBRL dataset, with the green line representing the data labels. There are point anomalies and contextual anomalies in the temperature data and correlation anomalies in the humidity data (positively correlated with temperature). The red line shows the detection results of Scheme 3, and the purple line shows the results of Scheme 4. Both schemes accurately detect point and contextual anomalies, but Scheme 4 almost fails to detect correlation anomalies. In contrast, Scheme 3, which employs the MFDGCN for extracting correlation information, effectively detects correlation anomalies.

These experimental results confirm the effectiveness of the proposed WSN anomaly detection model and validate that each component is essential, achieving excellent performance in WSN anomaly detection.

### E. Model Reliability Analysis

To verify the reliability of the designed model, we conducted multiple experiments and calculated the mean, variance, and standard deviation of the results to analyze the overall stability of the model. Specifically, the test set was divided into nine segments, seven of which were randomly selected for testing. This process was repeated 10 times, and the results are summarized in Table IV. In the table, Num represents the test iteration, Pre represents the precision, Rec represents the recall, and F1 represents the F1 score.

TABLE IV
Multiple Experimental Results

| Num | Pre (%) | Rec (%) | F1 (%) |
|---|---|---|---|
| 1 | 94.0 | 92.8 | 93.4 |
| 2 | 95.2 | 92.6 | 93.9 |
| 3 | 95.4 | 91.3 | 93.3 |
| 4 | 94.7 | 93.6 | 94.2 |
| 5 | 95.3 | 92.1 | 93.7 |
| 6 | 94.7 | 90.7 | 92.7 |
| 7 | 92.1 | 92.0 | 92.0 |
| 8 | 94.7 | 93.0 | 93.8 |
| 9 | 93.4 | 93.2 | 93.3 |
| 10 | 94.9 | 91.4 | 93.2 |

Table IV shows that over the 10 trials, the precision level ranged from a maximum of 95.4% to a minimum of 92.1%, and the recall ranged from a maximum of 93.6% to a minimum of 90.7%. These results indicate that the values of the three evaluation metrics were relatively stable. To further investigate the stability of the model and the reliability of its predictions, we calculated the mean, variance, and standard deviation of the results from these 10 experiments. The statistical results are presented in Table V, where "Metrics" denotes the evaluation indicators, "Mean" represents the average result, "Var" is the variance, and "Std" is the standard deviation.

TABLE V
Statistics of the Experimental Results

| Metrics | Mean | Var | Std |
|---|---|---|---|
| Pre (%) | 94.4 | 0.9 | 1.0 |
| Rec (%) | 92.3 | 0.8 | 0.9 |
| F1 (%) | 93.4 | 0.4 | 0.6 |

The mean values represent the overall performance of the designed model over the 10 experiments. As shown in Table 5, the mean values for the precision, recall, and F1 score metrics were 94.4%, 92.3%, and 93.4%, respectively. Although the mean values decreased slightly compared to the results of previous ablation studies (i.e., a 0.3% decrease in precision and a 0.1% decrease in the F1 score, with the recall remaining unchanged), the overall performance of the model remained high and showed good results. The variance and standard deviation reflect the stability and reliability of the model. From Table V, the standard deviations of the precision, recall, and F1 scores were 1.0, 0.9, and 0.6, respectively. Since none of the standard deviations exceeded 1.0, the designed model showed stable performance across the 10 experiments and consistently produced reliable results.

### F. Robustness Analysis

We also tested the sensitivity of the model to anomalous data by injecting noise at different levels to verify the robustness of the designed model. The method of injecting anomalous data is shown in Equation (28).

$$x' = x + \alpha(X^{max} - X^{min}) \qquad (28)$$

Where $x$ is the data of a specific sensor at a specific time of a randomly selected node in the dataset, and $X^{max}$ and $X^{min}$ represent the maximum and minimum values of the sensor data at that node, respectively. $\alpha$ is a set of coefficients that control the magnitude of the anomalous values, with values ranging from $\alpha \in \{-1, -0.5, -0.1, 0.1, 0.5, 1\}$. $x'$ is the data after injecting the anomaly.

The experimental results for different values of $\alpha$ are then summarized, as shown in Table VI. From Table VI, it can be seen that the three metrics, Precision, Recall, and F1 score, are all affected by the value of $\alpha$. They decrease as the absolute value of $\alpha$ decreases, which means that the effectiveness of the model in detecting anomalies is affected by the deviation of the anomalous data. The greater the deviation of the anomalous data, the easier it is for the model to detect the anomaly.

TABLE VI
Multiple Experimental Results

| $a$ | Pre (%) | Rec (%) | F1 (%) |
| --- | --- | --- | --- |
| -1 | 96.4 | 93.4 | 94.8 |
| -0.5 | 95.4 | 92.0 | 93.7 |
| -0.1 | 94.1 | 87.6 | 90.7 |
| 0.1 | 94.5 | 87.1 | 90.6 |
| 0.5 | 94.9 | 92.1 | 93.5 |
| 1 | 96.5 | 93.3 | 94.9 |

Although all three metrics decrease as the absolute value of $\alpha$ decreases, the effect on Precision is small, while Recall and F1 score are more significantly affected. From formula (27), it can be seen that the decrease in F1 score is primarily influenced by Recall. In other words, the deviation of anomalous data mainly affects the Recall metric of the model. According to Equation (26), Recall is mainly affected by True Positives (TP) and False Negatives (FN). Since Precision is almost unaffected by $\alpha$, we can conclude that the main reason for the decrease in Recall is the increase in FN. An increase in FN means that some anomalous data was not recognized by the model. This is because sensor measurements are easily affected by external disturbances and often contain noise, which causes some deviation between the measured data and the actual values. These deviations are usually not large and are not usually considered as anomalous data. Therefore, when there are anomalous data with smaller deviations, the model becomes less capable of recognizing these anomalies.

### G. Visualization Results and Analysis

In this section, we conduct a case study using a subset of the data and visualize the anomaly detection process of the proposed WSN anomaly detection method. Figure 7 shows the temperature data from node 21 in the IBRL dataset. The blue line represents the input data, which contain anomalies, while the orange line represents the data reconstructed by the model. Figure 8 displays the data labels indicating the locations of anomalies, with 0 representing normal data and 1 representing anomalous data. Figure 9 shows the computed anomaly scores and the selected threshold. The blue line represents the anomaly scores, while the red line represents the threshold. Figure 10 shows the anomaly detection results. Anomalies are identified by setting the data points with anomaly scores greater than the threshold to 1 and those with scores less than the threshold to 0. The specific anomaly detection results are shown in Figure 10. The anomaly detection results match the labels, indicating that the proposed method successfully identifies anomalous data.

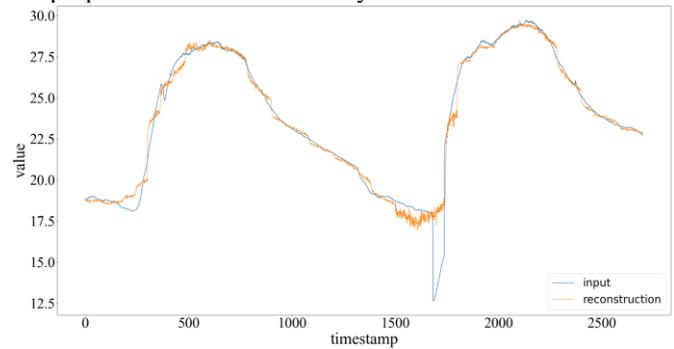

Fig. 7. Input Data and Reconstructed Data

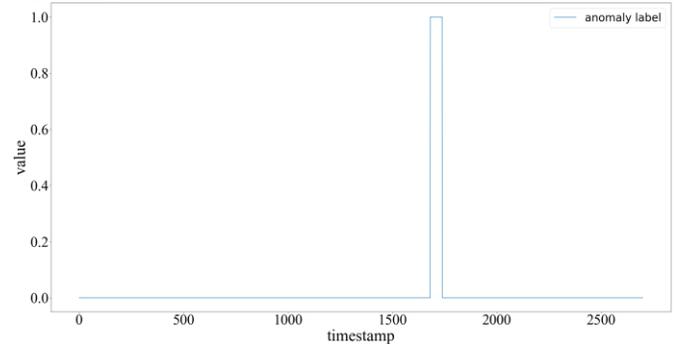

Fig. 8. Data Labels

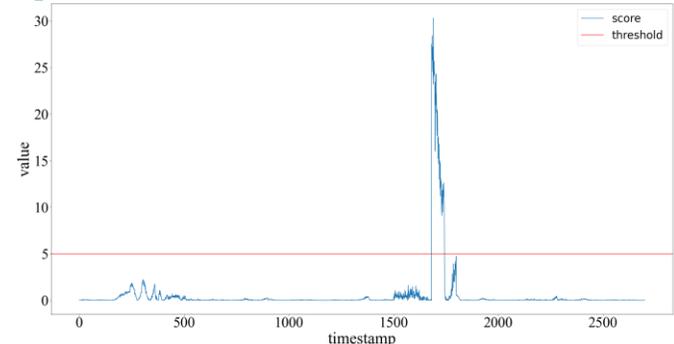

Fig. 9. Anomaly Scores and Threshold

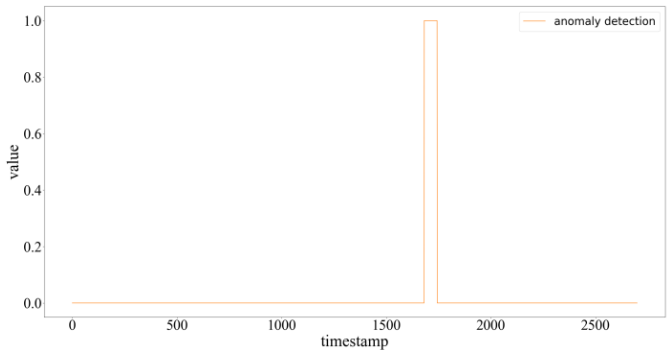

Fig. 10. Anomaly Detection Results

## VI. CONCLUSION

This paper proposes a WSN anomaly detection method that integrates frequency domain features and dynamic graph neural networks. The method addresses the limitations of existing transformer models in processing WSN data, especially in long-term dependency modeling, time-domain feature extraction, and spatial correlation feature extraction. The proposed approach uses a discrete wavelet transform to decompose the WSN data time series into trend and seasonal components, solving the long-term dependency modeling problem using a divide-and-conquer approach. This method effectively decomposes complex temporal patterns, enabling the model to more efficiently extract temporal features. Compared with the existing decomposition methods based on moving averages, it significantly reduces the incurred computational complexity. Given that WSN data typically exhibit strong periodicity and contain rich frequency components and that normal and anomalous data show significant differences in the frequency domain, anomalies can be easily distinguished in this domain. Therefore, the frequency domain attention mechanism is designed to fully utilize the features of data in the frequency domain, improving the accuracy of anomaly detection. Additionally, a multimodal fusion dynamic graph convolution module is introduced to enhance the model's ability to capture complex correlations between nodes. This module adaptively adjusts the network structure through an attention mechanism and integrates different modal features via cross-attention mechanisms. This allows the model to fuse features across different modalities and dynamically model spatial correlations. Experimental results demonstrate the effectiveness of the proposed method, which significantly outperforms baseline models in WSN anomaly detection tasks. However, the proposed method has the following limitations. First, although the DWT reduces the computational complexity of the model, the extensive use of attention mechanisms, which have quadratic computational complexity, means that the complexity of the model is not fundamentally reduced, making it challenging to handle large-scale WSN data. Second, the model is designed for data with certain periodic characteristics, so it may require further adjustments and optimizations for specific types of data, such as nonperiodic or highly nonlinear data. Future research will explore further feature extraction in the frequency domain, attempt to combine other graph learning methods like MPNN ,GAT and GraphSAGE, and integrate other deep learning techniques, such as contrastive learning, to further improve the anomaly detection performance. Additionally, new models that can fundamentally and effectively reduce the computational complexity of attention mechanisms will be designed to further improve the performance of the proposed model.

### ACKNOWLEDGMENT

The authors would like to thank the anonymous reviewers for their useful comments. This work is funded in part by the National Natural Science Foundation of China (Nos. 61861013, 62161006); Guangxi Innovation-Driven Development Project (Major science and technology project No.AA18118031); Guangxi Natural Science Foundation of China (No. 2018GXNSFAA050028); Director Fund project of Key Laboratory of Cognitive Radio and Information Processing of Ministry of Education (No. CRKL190102); Innovation Project of Guangxi Graduate Education(No.YCBZ2023134).

# Data Analysis

Before the dataset was used, we performed extensive analysis and preprocessing work. First, we discovered that the data derived from nodes 5 and 15 in the dataset had a significant amount of missing information. Figure 1 shows the data files for each node, and from the figure, it is evident that the files for nodes 5 and 15 were significantly smaller than those of other nodes. Therefore, we decided to exclude the data from these two nodes and only used the data acquired from the remaining 52 sensor nodes. Next, we found that the illumination data contained in the dataset were extremely poor, providing almost no useful information. Figure 2 shows the specific data for node 9, where panel (a) shows the illumination data (green curve), and panel (b) shows the data with the illumination data removed. As a result, we removed the illumination data and considered only the data derived from the humidity, temperature, and voltage modalities. We also observed several unreasonable data points. For example, the indoor temperature (blue line) should never exceed 120°C, and the humidity (orange line) should never be negative. We removed these unreasonable data points from the dataset. Finally, we found that different measurement attributes (modalities) often have different units, leading to significant magnitude differences between modalities. Figure 3 shows the data before and after performing Z score normalization (the blue line represents the temperature, the orange line represents the humidity, and the green line represents the voltage). In panel (a), before conducting normalization, the voltage data (green line) were barely visible, whereas in panel (b), after normalization, the changes exhibited by all three modalities were clearly visible. If we directly used the raw data for analysis purposes, the modality with the highest values would dominate the analysis, whereas the modality with the lowest values would be underrepresented. Therefore, to eliminate the influences of magnitude and value range differences between the modalities, we standardized the raw data to bring all modalities to the same scale. In this work, we used Z score normalization to standardize the input tensor, as shown in the following formula:

$$\mathbf{X}_{ij} = \frac{\mathbf{X}_{ij} - \mu(\mathbf{X}_{ij})}{\sigma(\mathbf{X}_{ij})} \tag{1}$$

where $\mathbf{X} \in \mathbb{R}^{N \times M \times W}$ represents the input tensor, $\mathbf{X}_{ij}$ represents the time series data measured by the $i$-th sensor at node $i$, $\mu(\mathbf{X}_{ij})$ is the mean, and $\sigma(\mathbf{X}_{ij})$ is the standard deviation. After performing Z score preprocessing, the modality time series data followed a normal distribution with a mean of 0 and a standard deviation of 1.

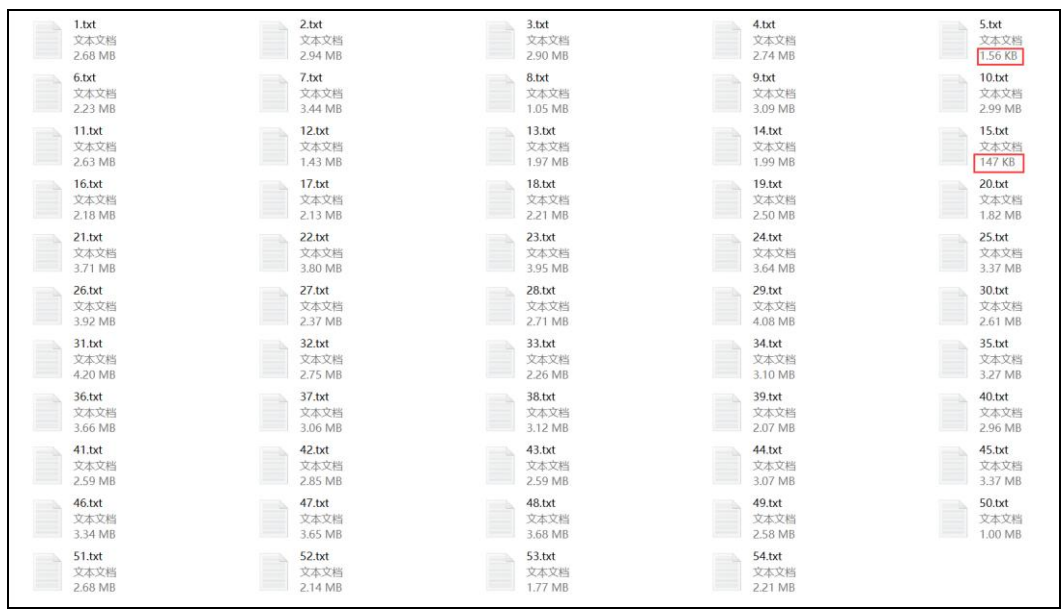

Fig. 1. Data files for each node

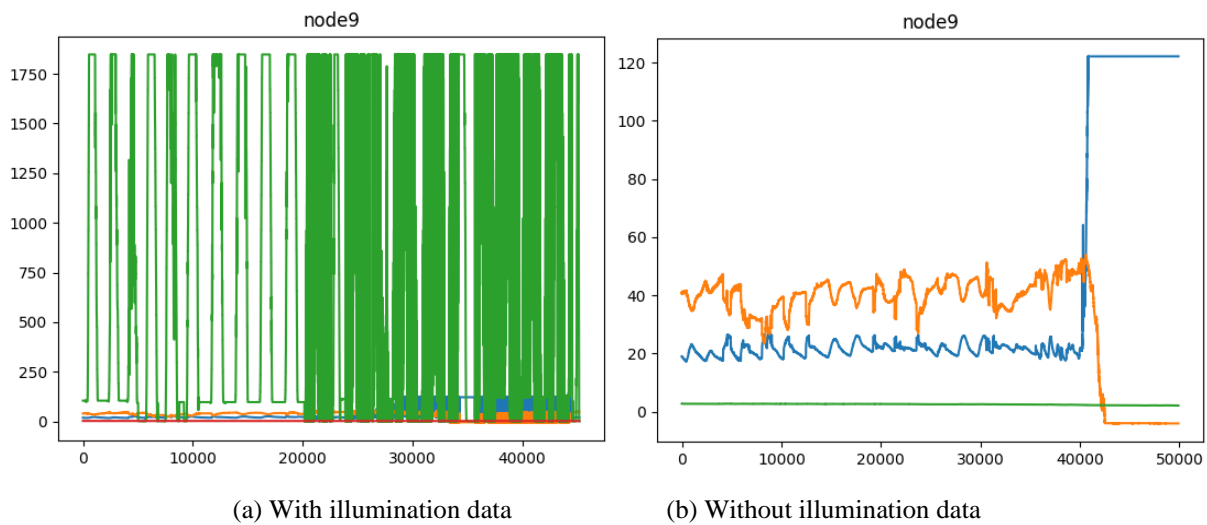

(a) With illumination data  (b) Without illumination data

Fig. 2. Sensor data for node 9

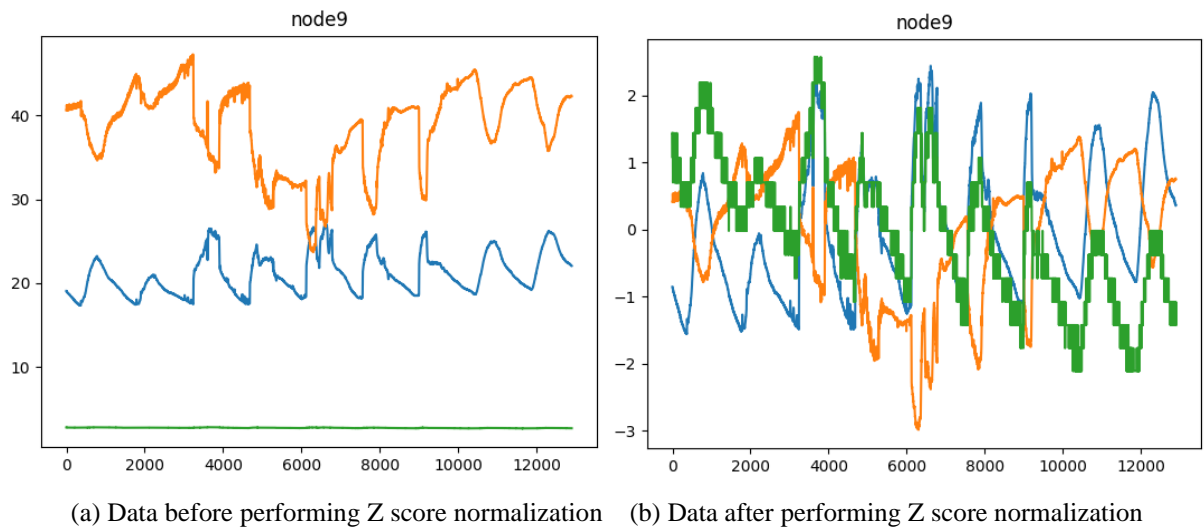

(a) Data before performing Z score normalization  (b) Data after performing Z score normalization

Fig. 3. Data before and after performing Z score normalization